# Rhodium Mössbauer Superradiance of Observable Gravitational Effect

Y Cheng and B Xia

Department of Engineering Physics, Tsinghua University, Beijing 100084, P. R. China

Email: yao@tsinghua.edu.cn

**Abstract.** We summarize the experimental observations of three case studies on the long-lived rhodium Mössbauer Effect. Extraordinary observations reported in this work manifest the open-up of photonic band gap in analogy to the superconducting gap. Observable gravitational effect is manifested by the superradiance of different sample orientations corresponding to the earth gravity. These observations are of potential importance for detecting gravitational waves and development of the two-photon gamma laser.

1. Introduction

Recently, we have reported the observation of long-lived rhodium Mössbauer Effect generated by bremsstrahlung irradiation [1]. We report several discoveries [2-4] in this work, i.e. (i) phase transition of nuclear exciton states at room temperature, (ii) spontaneous cascade down-conversion (SCDC) of entangled biphoton, (iii) SCDC energy gap, (iv) global photonic state, (v) macroscopic nuclear polarization, (vi) exciton diffusion enhanced by cooling, (vii) superradiant and subradiant nuclear Borrmann modes enhanced by cooling, (viii) bunching photons induced by the higher-order SCDC entanglement, (ix) non-Markovian photonic reservoir [5], (x) superradiance corresponding to the earth gravity.

The rhodium nuclear exciton as the photon-nucleus-photon bound state is a critical phenomenon depending on (i) inversion density, (ii) temperature, (iii) earth gravity, (iv) sample geometry and (v) magnetic field [2-4]. The former four effects are reported in the three case studies of this work. The dependence on magnetic field at room temperature will be reported elsewhere. In the case-I study [2], several exciton phases are discovered at room temperature. SCDC is identified to be the radiative branching channel of catalyzing the long-lived Mössbauer decay. In the case-II study [3], K suppression and γ suppression during the cooling period together reveal that the superradiance of SCDC branching are switched on, but not emitted in the direction toward the detector. In the case-III study [4], the superradiant direction is configured toward the detector. As responses to the liquid-nitrogen quenching, several phenomena are attributed to the direct observation of SCDC superradiance. Exciton diffusion and gravitational effect are demonstrated by arranging different positions of irradiated spot and different sample orientations corresponding to the earth gravity.

New physics emerges from the observations mentioned above. Temperature-dependent superradiance and exciton diffusion reveal the nuclear coupling as a function of temperature, *i.e.* long-lived Mössbauer Effect. SCDC entanglement is analogous to the entangled electrons of Cooper pair in superconductor. Natural linewidth of the SCDC γs is preserved by the spontaneous open-up of photonic band gap (PBG) [5-7] to inhibit the thermal agitation on photon energy, which is analogous to the superconducting gap. PBG phenomena of quantum optics related to our study such as localized superradiance, non-Markovian decay, photon-atom bound state, and anomalous Lamb shift, are depending on radiative transition [5], and radiator position [7]. Spontaneous PGB open leads to the Bose-Einstein condensation of exciton as the photon-nucleus-photon bound states with macroscopic polarization at room temperature. Entangled SCDC pair is bounded near nuclei in the most significant nuclear Borrmann channels (NBCs) of the low dimensionality. Exciton states pinched at defects are analogous to the color centers of semiconductor and the pinning center in superconductor. The global photonic state is revealed by two observations of emissions from polycrystal, i.e. macroscopic angular distribution and geometric dependence on sample size.

Superradiance can be driven by the gravitational waves [8], which opens up new approaches of compact detector for the gravitational waves particularly in the low frequency band. NBC is one of the focuses to achieve low-threshold γ laser [9,10]. SCDC defect mode in PBG materials [5,6] opens up a new approach for two-photon [11] γ laser.

**2. Origin of Rhodium Mössbauer Effect**

Rhodium has only one isotope $^{103}$Rh of 100% abundance in nature. The transition from $^{103m}$Rh to ground state is an E3 multipolar transition of the 56-minute half life. Single isolated rhodium nucleus will not give the observations of interest, which are thus attributed to the collective effect of identical rhodium nuclei in crystal.

Borrmann mode is the known x-ray eigenmode in crystal from field cancellation at the lattice sites such that the photo-electric effect is suppressed [12]. In the mean time, nuclear coupling can be enhanced for the multipolar nuclear transition [9]. Matching condition of the crystal lattice in the weakly attenuated NBCs provides the cooperative emissions of multiple γs with correlated directionality, polarization, and phase. The cooperative emission not able to be factorized into single-particle states [13] is similar to the entangled biphoton state generated by spontaneous parametric down conversion [13,14]. Since the Borrmann effect depends on temperature, it becomes more significant at low temperature due to the reduced atomic vibration. When atoms deviate from their equilibrium positions, they become scattering centers of entangled γs that leads to bunching arrival of X-rays. Consequently, the impurities, *e.g.* bismuth and platinum, are also scattering centers of the entangled γs.

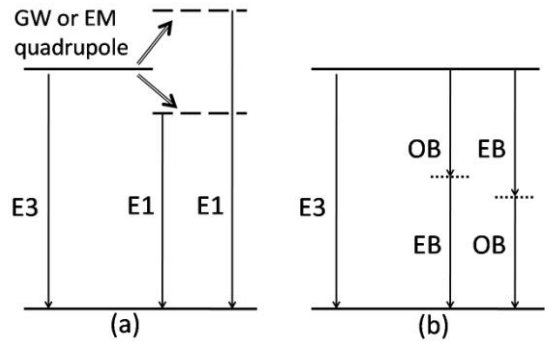

**Figure 1.** Schemes to catalyze the E3 multipolar transition. Dashed lines represent the virtual states. Dot line is the floating intermediate states of SCDC transition [15]. Single-line arrows are the spontaneous emission, whereas double-line arrows are external driving field. (a) The E3 multipolar transition becomes dipole transitions induced by gravitational Raman Effect [8] or by thermal electro-magnetic quadrupole field [15]. (b) The SCDC speed-up scheme. EB and OB are mNBMs of even parity and odd parity respectively.

SCDC is analogous to the spontaneous parametric down-conversion to generate entangled γs [13,14]. The scheme of figure 1a has been proposed to measure GWs [8], while figure 1b is suggested to be the reported SCDC scheme. Enaki and his collaborators theoretically investigated the SCDC cooperative emission of dipole-forbidden transitions among N radiators [15]. The collective down-conversion rate enters the regime of biphoton superradiance. Frequencies $\omega_1$ and $\omega_2$ of two paired photons are in general different but peaked at $\omega_0/2$. The observed SCDC spectral profiles peaked near $\omega_0/2$ manifest that the scheme of two-photon cascade illustrated in figure 1b has the major contribution. Two cascade photons ought to be the m-beam nuclear Borrmann modes (mNBMs) of similar spectral functions but opposite parity. OB wave and EB wave shown in figure 1b are the mNBMs of crystal symmetry such as $|3\sigma>_6$ and $|6\sigma>_6$ [9] toward the (1,1,1) direction respectively.

Conventional Mössbauer nuclei emit γs rarely into the 3D NBCs of a collimation angle typically on the order of $10^{-4}$ rad [9]. The situation drastically changes for SCDC branching channels, since SCDC wavevectors of varying energies freely match the NBCs. The PBG of dielectric photonic crystal is fixed [6], whereas the PBGs of NBCs are dynamic in analogy to the superconducting gap. The SCDC scattering in NBCs has complete suppression of all elastic, inelastic resonant processes, and coherently enhanced by N×m for N nuclei and mNBMs [9]. SCDC localization occurs in the rhodium NBCs, of which the mean free paths approach the lattice constant of 0.38 nm for N×m∼30. When PBG spontaneously opens up, localized SCDC superradiance emits into the most significant 1D reservoir of NBCs [5] and *vice versa*. Though the John's $N^3$ superradiance [5] and the Hannon-Trammell anomalous emission [9] are different in many aspects, they share the same effect that all the isotropic emissions in 4π-sr vanish for sufficiently large N. Infinitesimal impurity defects or grain boundaries in polycrystal lower the threshold of phase transition. Observed PBG in this report is about 200 eV due mainly to the resolution of detecting system, which requires further investigation. In analogy with the exciton of electron-hole pairs bounded by Coulomb force in semiconductors, SCDC γs are bounded by the nuclear susceptibility as the photonic potential well [6]. Moreover, in analogy to the entangled electrons in superconductors of Copper pair created by electron-phonon-electron interaction, SCDC entanglement is suggested by the photon-polariton-photon interaction, where polariton is the collective nuclear polarization in resonance with mNBMs guided in NBCs.

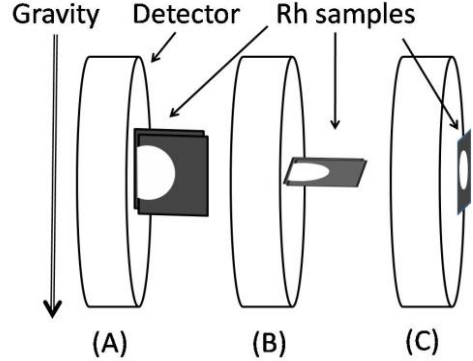

**Figure 2.** Sample orientations. Bright spots of A and B indicate the irradiation spot located near the sample edge. Irradiation spot of C is at the sample center. (A) Standing orientation in the case III. (B) Leveled orientation in the case III. (C) This orientation is applied in the case I and the case II. Bremsstrahlung irradiation is carried out in C by replacing detector with accelerator.

## 3. Experimental set-ups

In this work, we show three case studies of applying two different measuring systems and three different sample orientations (figures 2A-2C). Case I demonstrates the exciton phase transitions at room temperature. Case II demonstrates that the superradiance is switched on by cooling. Case III demonstrates the direct observation of superradiance and its associated gravitational effect. The square sample has a dimension of 2.5 cm × 2.5 cm × 1 mm with 99.9% purity of rhodium (Goodfellow Rh00300). The rhodium sample is a polycrystal with the fcc lattice structure.

*3.1. Case I: sample orientation C at room temperature*

We have applied 120-minute bremsstrahlung irradiation of the same excitation procedure reported in [1]. The CANBERRA detection system consists of a HPGe low-energy detector GL0510P, an optic feedback pre-amplifier (CANBERRA 2008 BSL) and a digital multichannel analyzer (CANBERRA DSA 1000). The active detecting area is 500 mm$^2$ covering an emitting solid angle about $1.2\pi$ sr oriented in the NS direction. Measurements are carried out with the horizontally leveled HPGe detector illustrated in figure 2C. Bremsstrahlung intensity applied in the case I is fivefold stronger than the intensity applied in the case II and in the case III.

*3.2. Case II: sample orientation C with quenching*

The sample orientation is the same as described in case I (figure 2C). A small HPGe detector of CANBERRA GL0210P with an active area of ϕ-1.6 cm is applied in the N-S detecting direction. Collecting angle is about $0.4\pi$ sr. The detecting system consists of an optic feedback pre-amplifier (CANBERRA 2008 BSL), a multichannel analyzer (ORTEC 917A) and an ORTEC 572 amplifier tuned at 3-µs shaping time. The sample is immersed by liquid nitrogen from one side. Typical temperature behavior is measured later as shown in [3].

*3.3 Case III: sample orientations A and B with quenching*

Detecting system and irradiation intensity are the same as the case II, except for the different sample preparation. Two 1mm×25mm×25mm rhodium samples sandwiched by 0.5-mm tungsten and 0.05-mm plastic sheets are stacked together such that the 25-mm direction is toward the detector. Sample orientations are illustrated in figures 2A and 2B by the orientation A (denoted by A) and the orientation B (denoted by B). Transient response of liquid-nitrogen quenching becomes several minutes due to the additional heat capacity and the thermal isolation.

## 4. Observations

*4.1. Case I: sample orientation C at room temperature*

Three measurements of inserting filters are shown in this case. Filters are: (a) none; (b) 35-µm copper; (c) 25-µm tantalum. The initial count rate is about 12000 cps mainly contributed by K lines in (a). Figures 3a-3c illustrate the time evolution of spectral deformations $\tilde{S}_i(\omega,t)$ deviated from the normal profiles $\bar{S}_i(\omega)$. Measured spectra $S_i(\omega,t)$ are normalized by individual total counts, channel by channel in time

$$\tilde{S}_i(\omega,t) = A_i \left( \frac{S_i(\omega,t)}{\int S_i(\omega,t)d\omega} - \bar{S}_i(\omega) \right) \tag{1}$$

with factors $A_i$ chosen for each band for figure 3a-3c and figure 8. The subscript $i$ stands for Kα and Kβ and γ. Phase transitions are identified by satellites and hypersatellites of K lines [16]. Preliminary assignments of six phases are listed in Table 1. Several phases are caused by the filter absorption such as H3 and part of H5 may be the same H1 phase, which will be reported elsewhere. Of particular concern is the revival of H1 phase in figure 3b, which has been observed in many other experiments. In the cavity QED [13,17], revival manifests the strong coupling. The reported experiments are selected to demonstrate the phase variety. To compare spectral profiles in details, we average the spectral deformations of each phase in figures 4a-4c. The spontaneous phase transition is shown by figure 4d. Energy distributions of γ at 39.76 keV shown in figures 4a-4c manifest the PBG.

**Table 1.** Preliminary assignments of exciton phases in the case I. Some phases are probably the mixed states of elementary modes.

| Phase | H1 | H2 | H3 | H4 | H5 | L1 |
|---|---|---|---|---|---|---|
| Time (ks) | (a) 0-2.8, (b) 8-9.3 | (a) 5.2-11 | (b) 0-2.1 | (b) 2.1-8 | (c) 0-8.8 | (a) 2.8-5.2 |

The deformations of Kα and Kβ spectra are mainly attributed to satellites and hypersatellites of multiple ionizations [16]. There are two typical deformations in the Kα band, i.e. energy shift and enhancement of $K\alpha_2$. Energy shifts of rhodium K satellites and K hypersatellites are about 50 eV and 500 eV due to coincident creation of two holes, i.e. one K hole + one L hole and two K holes respectively [16]. Three $L_1$ holes are possible, whereas three $L_2$ holes are impossible due to the fact that $L_2$ shell has only two electrons. In the case of two K holes + one L hole, Kα hypersatellite is emitted first, which gives states of two L holes + one K hole. When subsequent second photon emits, the possibility to emit $K\alpha_1$ satellite is higher than the possibility to emit $K\alpha_2$ satellite. Thus, $K\alpha_2$ count is enhanced by the normalization of total Kα band. Observed Kβ hypersatellites are generally stronger than Kα hypersatellites, which indicate the coincident creation of two K holes + multiple L holes. Multiple ionizations on the same rhodium nucleus analyzed above manifest the higher-order entanglement of SCDC γs.

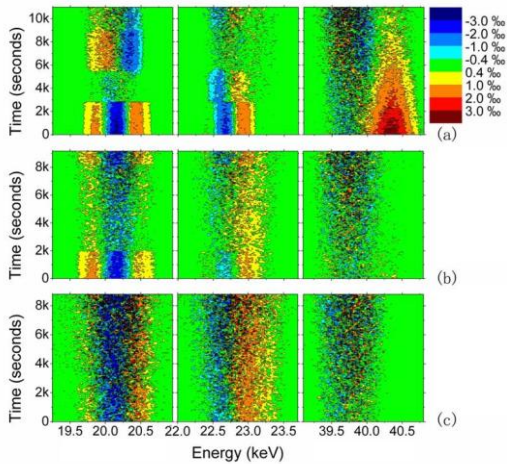

**Figure 3.** (Color online) Time evolution of spectral deformations, i.e. Kα lines at 20 keV, Kβ lines at 23 keV, and γ at 40 keV in the case I. Total counts $N(\omega,t)$ in each band are normalized with $N(K\alpha,t)=1$, $N(K\beta,t)=0.5$ and $N(\gamma,t)=0.25$ for clear presentation as defined in eq. (1). Absorption sheets suppress the Kα peak pile-up located at right shoulder of the 40-keV γ peak. Filters are: (a) none; (b) 35-μm copper; (c) 25-μm tantalum. Preliminary phase assignments are listed in Table 1.

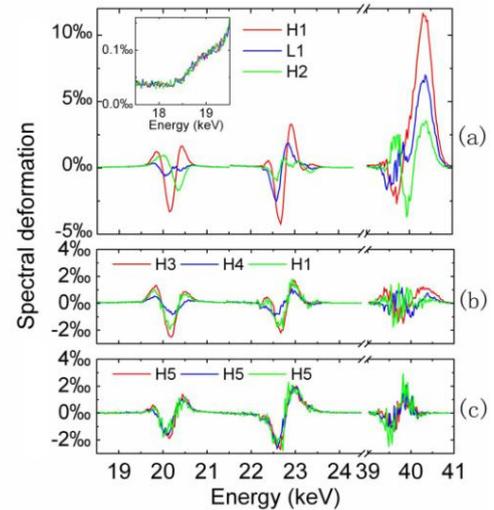

**Figure 4.** (Color online) Averaged spectral deformations in the case I corresponding to each phase illustrated in figures 3a-3c except all $N(\omega,t)=1$. Ordinate scales are same for three figures. SCDC γs near 20 keV are illustrated in the blow-up figure. Time sections of 0-2500s-6100s-8800s of figure 4c are arbitrarily selected.

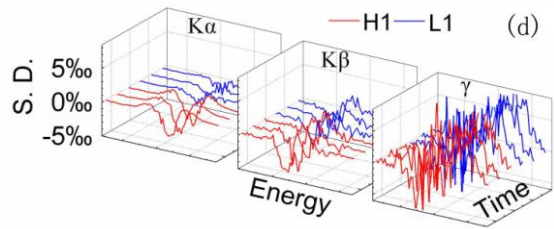

**Figure 4d.** (Color online) The original data per minute of figure 4a to demonstrate the prompt phase transition.

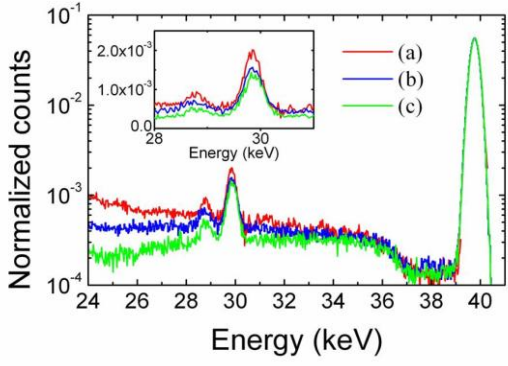 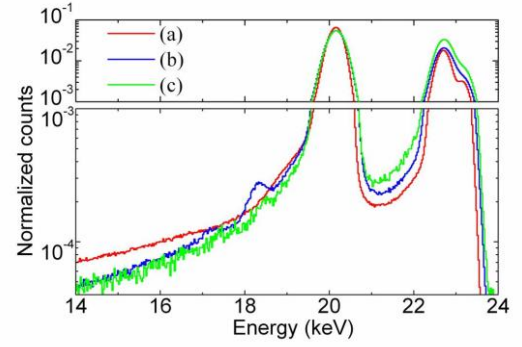

**Figure 5.** (Color online) SCDC spectral profile of $c\gamma2$ between 24 keV and 38 keV demonstrated by filters, i.e. (a) none; (b) 35-μm copper; (c) 25-μm tantalum in the case I. We have removed residual pile up by calculation. The Compton continuum ($<10^{-4}$) is negligible. SCDC counts in the order of $10^{-3}$ are normalized by measured γ counts at 40 keV. The net counts of main escapes peak at 30 keV shown in the blow-up figure excluding the pile-up contribution are (a) 0.023 (b) 0.019 (c) 0.018. Peaks are slightly broadened (<10%). The number 0.023 of (a) is 60% stronger than our calibration.

**Figure 6.** (Color online) SCDC spectral profile between 14 keV and 24 keV demonstrated by filters, i.e. (a) none; (b) 35-μm copper; (c) 25-μm tantalum in the case I. Spectra are normalized by the Kα. Sidebands near 17.2 keV and 18.3 keV indicate the higher transmission of entangled photons than that of individual photons. Slight peak shifts between green and blue lines are observable. Trapezoid shaping times of DSA multichannel analyzer are (a) 5μs rise time, 0.8μs flat top; (b) and (c) 2μs rise time, 0.6μs flat top, which lead to slightly narrower FWHM of (a) than FWHM of (b) and (c).

SCDC γs shown in figures 5 and 6 follow the γ decay rate in time as the branching decay. Two spectra of energy > 24 keV (denoted by $c\gamma2$) and < 24 keV (denoted by $c\gamma1$) resolved by filters indicate the SCDC peak near 20 keV (see figures 4a, 5, 6, and 10a-10c). Total counts between 24 keV and 38 keV normalized by γ count ($c\gamma2/\gamma$) is 0.297±0.001 (~1.2π sr in the case I), 0.452±0.003 (~1.7π sr, figure 2b in [1]), 0.446±0.003 (~1.7π sr, figure 2a in [1]), 0.358±0.002 (~0.4π sr in the case II), 0.36±0.05 (~0.4π sr in the case III) at room temperature. The observed deviations are due to the different collecting solid angles. It becomes 0.59±0.03 in the case III at low temperature. Except the case of figure 2b in [1] is measured by the EW oriented detector, all the other cases are measured by the NS oriented detector. Thus, SCDC γs emitted from the polycrystalline sample have macroscopic angular distribution. In figure 6, two LHS sidebands in $c\gamma1$ are identified at 17.4 keV and at 18.3 keV, respectively. Two sidebands are slightly different in the two filtered cases, which requires further investigation. Total SCDC counts are nearly equal to the γ counts at 40 keV. We demonstrate the SCDC entanglement by the γ escape at 30 keV in addition to hypersatellites. Escape from the detector plays a role of the Hanbury-Brown-Twiss interferometer [17] to manifest the multiple-photon entanglement. When several entangled γs of fixed total energy arrive at the detector together, individual low- energy γs give the high escape probability. The γ escape in the blow-up figure of figure 5 reveals that SCDC γs are entangled.

*4.2. Case II: sample orientation C with quenching*

The initial count rate is about 700 cps mainly contributed by K lines. Figure 7 illustrates the time evolution of the two K lines, γ and γ escape. The speed-up decay during the cooling period demonstrates the on-switch of superradiant channels, while the immediate recovery after quenching stop demonstrates the off-switch of superradiant channels.

Since the superradiant direction is oriented along the long edge (25-mm direction) of the polycrystalline sample, the SCDC subradiance toward the detector is revealed by the sag of γ escape shown in the blow-up figure of figure 7.

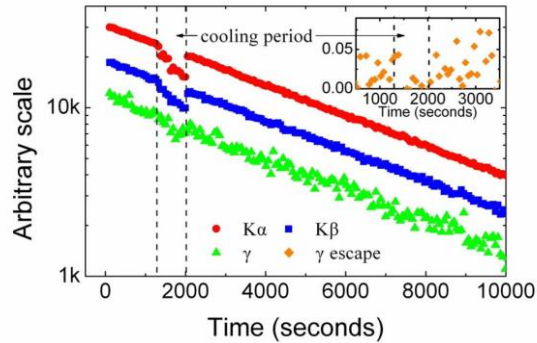

**Figure 7.** (Color online) Time evolutions of two K lines, γ and γ escape in the case II. The ordinate scale stands for the Kα but arbitrary for Kβ and γ. Superradiance channels are switched on but not toward the detector, which is revealed by the recovery back to original decay lines immediately after cooling stop. Suppression of K lines and γ are attributed to directional SCDC superradiance. Time evolution of 30-keV γ escape normalized by 40-keV γ is shown in the blow-up figure. Sag of γ escape induced by cooling reveals the subradiance of entangled SCDC γs.

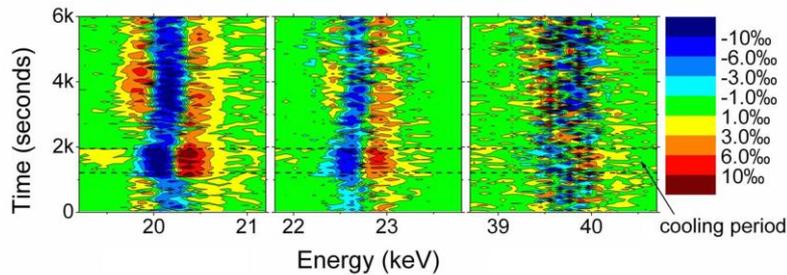

**Figure 8.** (Color online) Time evolutions of three spectral deformations, i.e. Kα lines at 20 keV, Kβ lines at 23 keV, and γ at 40 keV in the case II. Total counts N(ω,t) in each band are normalized with N(Kα,t)=1, N(Kβ,t)=0.5 and N(γ,t)=0.25 for clear presentation as defined in eq. (1). The baseline shift in the original data has been corrected. Hypersatellites are increased by an order of magnitude after quenching. Normalizations in each band are the same values in figures 3a-3c. $Kα_2$ is enhanced except during the cooling period, in which the satellite shift dominates. After cooling stops, hypersatellite shift increases in addition to satellite shift, where Kβ hypersatellites are stronger than Kα hypersatellites.

Figure 8 explores the time evolution of spectral deformations defined in (1). K satellites and K hypersatellites are enhanced by lowering the sample temperature. During the cooling period, immediate satellite shifts with delayed hypersatellite shifts are found. Satellite shifts are revealed by zero crossing lines of eq. (1) locating at 20.2 keV of Kα and at 22.7 keV of Kβ respectively. When temperature slowly recovers back to room temperature, spectral deformation becomes similar to the H1 phase in the case I featured by significant hypersatellites. On the left hand side of the Kα around 19 keV in figure 8, the SCDC peak emerges during cooling period, which shall be better demonstrated in the case III.

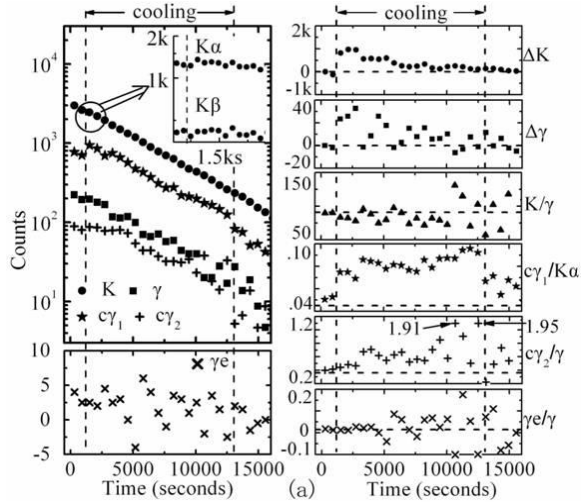
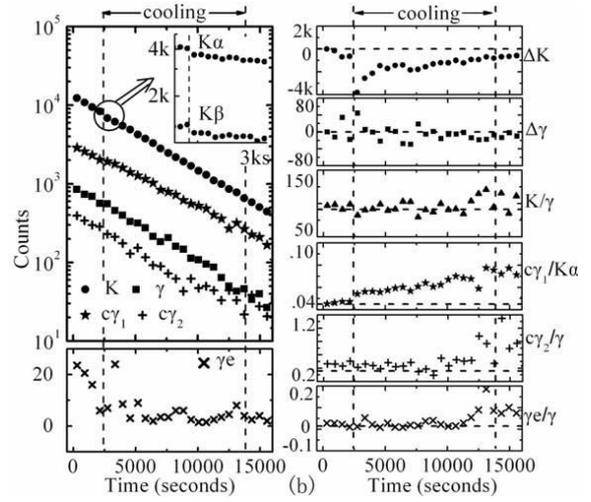

**Figure 9.** Time evolution of K lines, γ, γ escape, and SCDC γs in two bands of cγ1 and cγ2. Abbreviations are the same in Table 2. The LHS ordinate scale stands for γ, cγ1, and cγ2, whereas K scale is arbitrary. Blow-up figures of separated Kα and Kβ lines after cooling are explored by original data measured per minute. All the other data are accumulated in ten minutes. The derived values are illustrated on the RHS, in which the dashed lines give the calibrated values listed in Table 2b. ΔK and Δγ are the deviations to the normal decay given by eq. (2). The calibrated cγ1/Kα level of 0.040 is mainly (70%) contributed by the detector imperfection from Kα.

**Figure 9a.** *Central irradiation spot in orientation A.* Kβ sag instead of Kα jump shown in the blow-up figure reveal the K suppression and 5% contribution from SCDC count. The increment of cγ1/Kα reveals that K increment is not attributed to cγ1 at 20 keV but exciton diffusion.

**Figure 9b.** *Edge irradiation spot in orientation A.* In the blow-up figure, the Kβ sag is deeper than the Kα sag attributed to K suppression and SCDC contribution respectively. K suppression is more significant than γ suppression shown by K/γ ratio. Ratios of γe/γ reveals the entanglement developed in time.

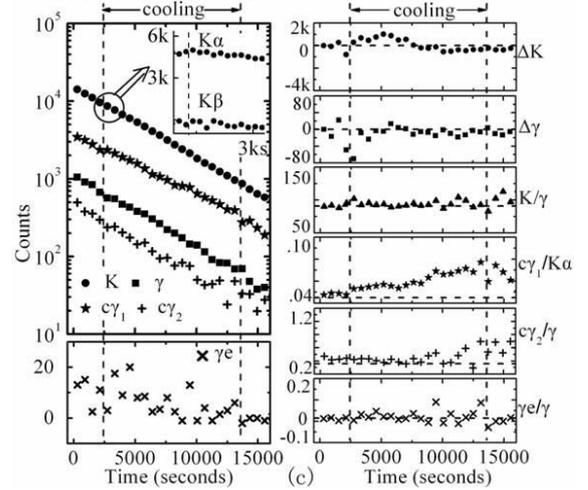

**Figure 9c.** *Edge irradiation spot in orientation B.* The data point before cooling is strange [19]. Superradiance and K suppression become insignificant in B. The exciton diffusion is firstly increased and then decreased due to the fact that the irradiation spot center locates not exactly at edge as shown in figure 2B. This decentralization is enhanced due to 2D diffusion path in B instead of 1D diffusion path in A. Ratios γe/γ also reveal weaker entanglement than that observed in figure 9b.

*4.3. Case III: sample orientations A and B with quenching*

Experimental conditions and relevant observables are listed in Tables 2. Time evolution of characteristic emissions from rhodium sample is illustrated on LHS of figures 9a-9c. On the RHS, details are explored by two derived values, i.e. ratios and deviations. Deviations from the normal decay are given by

$$\Delta N_{K,\gamma} = N_{K,\gamma}(t) - N_{K,\gamma}(0)e^{-t/\tau}, \qquad (2)$$

where $N_{K,\gamma}$ are counts of K lines and γ. $\tau$ is the natural $^{103m}$Rh lifetime of 4857s. Jump of cγ1 count immediately after cooling in figure 9a reveals the direct observation of SCDC superradiance. Exciton diffusion is revealed by moving the irradiation spot from the center to the edge. Slow diffusions toward and away from the detector are shown by the positive and the negative ΔK in figures 9a-9c, respectively. When the cooling is stopped, superradiant modes are switched off and leads to an immediate decrease of K decay rate.

**Table 2a.** Experimental conditions and relevant observables in the case III. Indices (a), (b), and (c) are corresponding to figures 9a-9c. Abbreviations of "K, cγ1, cγ2, γe, $N_K(0)$, $\alpha_K$, and $\Delta\tau_{K\beta}$." stand for K lines including Kα and Kβ, SCDC count of 14-19 keV, SCDC count of 24-38 keV, γ escape, initial count rate, internal conversion of K lines, and the rate change of K lines at cooling stop, respectively. Index (i) indicates the average value before the cooling start. Index (ii) indicates the average value during cooling. Index (iii) indicates the average value after cooling stop.

| No. | Spot | Orientation | (i) $\alpha_K$ | (ii) $\alpha_K$ | (iii) $\alpha_K$ | (i) cγ1/γ | (ii) cγ1/γ | (iii) cγ1/γ |
|---|---|---|---|---|---|---|---|---|
| (a) | center | A | 90.1±4.5 | 83.6±2.2 | 99.1±8.8 | 0.047±0.001 | 0.073±0.001 | 0.056±0.003 |
| (b) | Edge | A | 96.3±1.8 | 91.8±1.5 | 87.1±5.5 | 0.044±0.0005 | 0.058±0.0005 | 0.064±0.002 |
| (c) | Edge | B | 91.1±1.5 | 94.3±1.4 | 95.0±4.3 | 0.045±0.0004 | 0.053±0.0004 | 0.054±0.01 |

| No. | $N_K(0)$ | $\Delta\tau_{K\beta}$ | (i) cγ2/γ | (ii) cγ2/γ | (iii) cγ2/γ | (i) γe/γ | (ii) γe/γ | (iii) γe/γ |
|---|---|---|---|---|---|---|---|---|
| (a) | 37.4 cps | -94±55 s | 0.41±0.04 | 0.59±0.03 | 0.53±0.15 | 0.016±0.009 | 0.020±0.006 | 0.029±0.04 |
| (b) | 146 cps | -33±27 s | 0.46±0.02 | 0.43±0.05 | 0.56±0.06 | 0.023±0.005 | 0.027±0.004 | 0.064±0.02 |
| (c) | 178 cps | -33±24 s | 0.44±0.01 | 0.41±0.01 | 0.66±0.03 | 0.014±0.004 | 0.023±0.004 | -0.017±0.009 |

**Table 2b.** Calibrations of characteristic parameters in Table 1a without cooling. Four values of the C measurement are adopted to be the calibrated numbers shown in figures 9a-9c.

| Orientation | spot | $N_K(0)$ | $\alpha_K$ | cγ1/Kα | cγ2/γ | γe/γ |
|---|---|---|---|---|---|---|
| A | center | 38 cps | 92.8±1.7 | 0.0452±0.0005 | 0.36±0.05 | 0.014±0.005 |
| B | center | 44 cps | 90.1±1.4 | 0.0454±0.0005 | 0.33±0.04 | 0.022±0.005 |
| C | center | $3\times10^3$ cps | 91.3±0.2 | 0.040±6.4×10$^{-5}$ | 0.358±0.002 | 0.0196±0.0008 |

The SCDC spectrum is peaked near 20 keV with energy shift (figures 10a-10c). One sideband at 18.3 keV observed in the case I is missing. Slightly reductions of sidebands after cooling occur in figures 10a and 10b but not in figure 10c. When quenching starts in figure 9a, Kα jump and Kβ sag reveal the superradiance near 20 keV and slightly observed K suppression respectively. By moving the irradiated spot from sample center to the sample edge, we obtain Kα sag and Kβ sag in figure 9b. K suppression becomes significant in figure 9b, where Kβ sag is deeper than Kα sag due to SCDC jump contribution in Kα counts. The obtained γ suppression is always less significant than the K suppression due to the coincident detection of entangled SCDC γs. Increment of γ escape in the case III instead of sag in the case II reveals the entangled γs propagating toward detector in the case III.

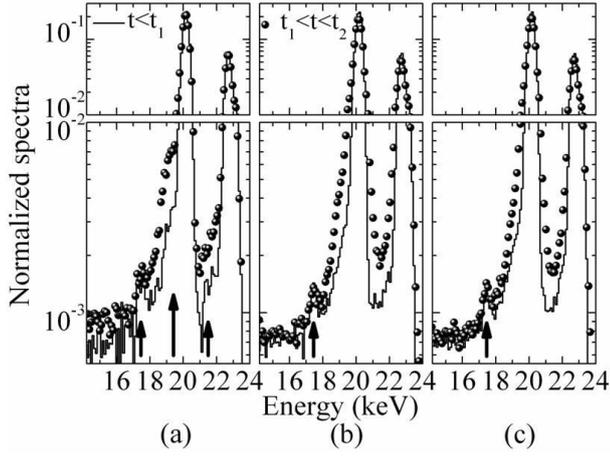

**Figure 10.** Cooling change of K spectral profiles normalized by Kα counts in the case III. Indices (a), (b), and (c) are corresponding to figures 9a-9c. Solid lines are measured before cooling, points are obtained after cooling. The quenching starts at $t_1$ and stops at $t_2$. Left sideband at 17.4 keV appears in all of three measurements. (a): One main peak at 19.4 keV and two sidebands at 17.4 keV and 21.5 keV, which are indicated by arrows. Slight reductions of peak at 17.4 keV occur after cooling in (a) and in (b) but not in (c).

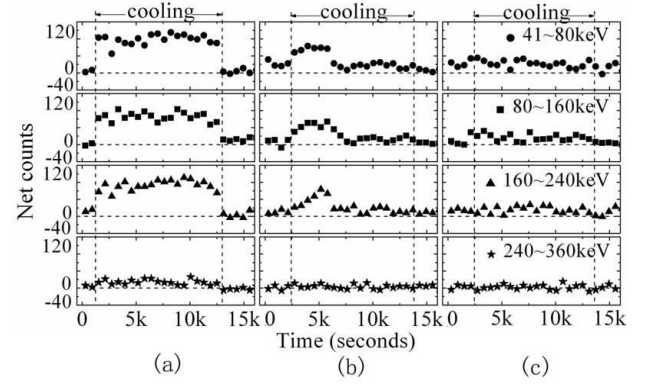

**Figure 11.** Bunching photons displayed in four different bands in the case III. Indices (a), (b), and (c) are corresponding to figures 9a-9c. Bunching photons transmitted in NBCs shall arrive at the detector with angular distribution. Thus, the earth gravity and the sample geometry collimate the photons in A but not in B. The data point of (c) before cooling is strange [19].

Gravitational redshift restricts the nuclear coherence in a stripline of 30 nm × 1 mm × 25 mm when sample is in A. The earth gravity traps γ of atto-eV natural linewidth in the vertical span of 30 nm [1], while the sample thickness is 1 mm. When the sample is rotated to B, gravitational restriction is extended to be the 2D slice of 30 nm × 25 mm × 25 mm. In the anisotropic 3D photonic crystal, the superradiant intensity of 1D-PBG system is stronger than that of the 2D-PBG system [5]. In the conventional sense, the K internal conversion rate $\alpha_K$ shall not depend on temperature and orientation. The $\alpha_K$ discrepancy between A and B (see figures 9a-9c and Table 2a) demonstrates the gravitational effect on superradiance, *i.e.* $\alpha_K$ decrease due to K suppression. Moreover, exciton diffusion and SCDC superradiance also reveal the gravitational effect between A and B (see figures 9a-9c and Table 2a). Figures 11a-11c illustrate the immediate quenching response of bunching photons to demonstrate the gravitational effect on the superradiance of multiple entanglement.

Figures 12a-12c illustrate the time evolution of spectral deformation defined by eq. (1). The deformations observed in A and in B are order-of-magnitude stronger than the deformations observed in C. The delayed onset of hypersatellite and its hysteresis effect is attributed to the non-Markovian reservoir [5].

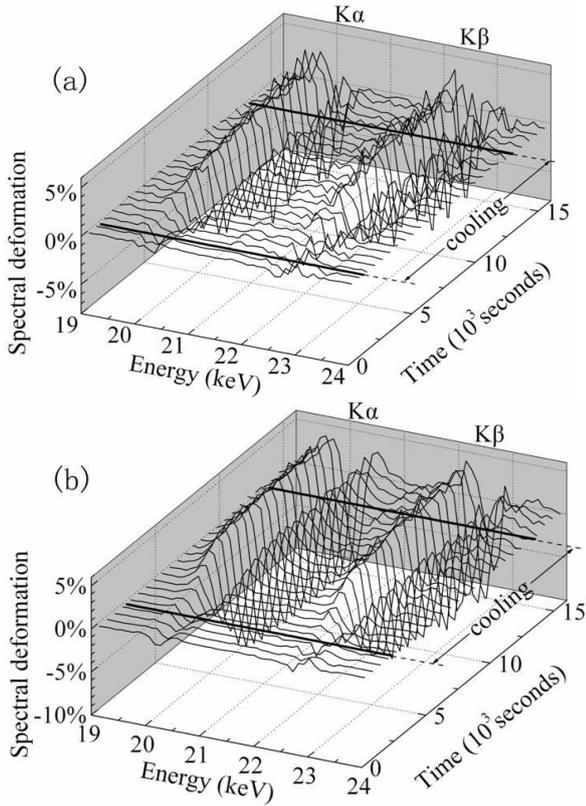
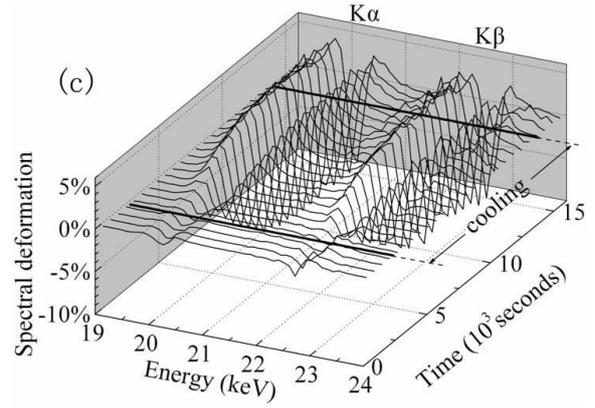

**Figure 12.** Normalized spectral deformations of K lines defined by eq. (1). Indices (a), (b), and (c) are corresponding to figures 9a-9c. Spectral deformation in A of (b) increases faster, stronger and stays longer than the deformation in B of (c).

## 5. Discussions

As responses to the liquid-nitrogen quenching, we observe plenty of phenomena. The observations in the case II are summarized as (i) immediate speed up of K lines and γ; (ii) immediate sag of γ escape; (iii) immediate jump of the bunching-photon count (not shown); (iv) immediate recovery of speed-up decay after quenching, (v) delayed on-set of hypersatellites. When we rotate the sample orientation from C to A in the case III, the responses to quenching become (i) immediate jump of the bunching-photon count; (ii) immediate jump of the SCDC count; (iii) immediate K suppression; (iv) delayed onset of hypersatellites; (v) hypersatellite hysteresis; (vi) delayed onset of the γ escape. Former three observations are attributed to direct observation of γ superradiance. Later three observations are attributed to the non-Markovian reservoir [5]. Observations of the case II and the case III demonstrate the superradiance depending on the macroscopic sample size. After irradiation, sample becomes a heat source. The quenching process generates significant temperature variation insides the sample. Observed superradiance reveals that the nuclear coupling is insensitive to the temperature distribution but sensitive to the earth gravity. The PBG open is thus manifested. Additional evidences are provided by the exciton phase transitions in the case I. PBG shall be higher than the thermal potential at room temperature. However, the overly large PBG about hundreds of eV is puzzling.

SCDC emit the energy-time entangled γs into the reservoir near 1D photonic band gap (PBG) [5], i.e. into particular NBCs of suppressed photo-electric attenuation, suppressed conversion, strong nuclear coupling [9], and confined coherence by the earth gravity and the temperature distribution in polycrystal. The SCDC spectrum peaked near the half transition energy (20 keV) manifests that the major contribution is the biphoton cascade. Because of the E3 multipolar transition, SCDC biphoton must be mNBMs of opposite parity. Collective spontaneous rates of two different mNBMs depend on PBG environment. SCDC spontaneous rate of coupled nuclei enters the regime of biphoton superradiance, i.e. proportional to N of N nuclei [15] or $N^2$ of the John's superradiance in the most significant 1D NBCs [5]. Higher-order coherence of SCDC γs emitted from different nuclei developed in time. When measurement starts, the higher-order coherence has been developed for a 120-min irradiation time. Temperature sag immediately enhances the transmission of the existing mNBMs. Bunching photons of total energy ranging from 41 keV to 360 keV arrive at the detector together. When the sample is rotated from A to B, the significant reduction of bunching photons reveals that the 1D NBCs are extended to 2D NBCs ascribed to the earth gravity.

## 6. Conclusions

We directly observe the superradiance in the particular direction along the long edge of the polycrystalline sample. Speed-up decay and exciton diffusion manifest the long-lived Mössbauer Effect, though the Doppler shift [18] shall prohibit the nuclear resonance. Our experiments have shown different aspects that PBG inhibits the thermal agitation of photon energy. Spontaneous open of PGB leads to Bose-Einstein condensation of exciton as the photon-nucleus-photon bound states. Gravitational effects corresponding to different sample orientations demonstrate the nuclear coupling by γs of atto-eV natural linewidth.

SCDC spectral profile peaked near 20 keV reveals that the biphoton SCDC transitions have major contribution. Higher-order entanglement enhanced by quenching builds up in time. Branching SCDC γs emitted by the macroscopic angular distribution from polycrystal reveal a global photonic state. Geometric dependence of macroscopic sample size manifests the long-range coupling across grain boundaries of polycrystal despite the significant temperature variation inside the sample; otherwise superradiance is not only guided in the 25-mm direction but also 1-mm direction for polycrystal. Exciton phases are identified by the emissions from exciton states pinched at defects, which are analogous to the color centers of semiconductor and the pinning center in superconductor.

## 7. Acknoledgements

We thank Hong-Fei Wang and Yi-Kang Pu for the help of manuscript preparation, Yanhua Shi, Ye-Xi He and Yi-Nong Liu for fruitful discussion and Yu-Zheng Lin with his accelerator team of the department, especially Qing-Xiu Jin and Xiao-Kui Tao. YC would like to thank J. P. Hannon for his inspiring critical comments. YC appreciates the mental support of years by Yuen-Chung Liu in the National Tsing Hua University at Hsinchu, Taiwan. This work is supported

by the NSFC grant 10675068.

## 8. References

[1] Cheng Y, Xia B, Liu Y-N, Jin Q-X 2005 *Chin. Phys. Lett.* **23** 2530; Cheng Y, *et al.* 2006 *Hyperfine Interactions* **167** 833

[2] Cheng Y, Xia B 2007 ArXiv:0706.2628

[3] Cheng Y, Xia B, Wang Z-M 2007 ArXiv:0706.2620

[4] Cheng Y, Xia B 2007 ArXiv:0707.0960

[5] John S and Quang T 1995 *Phys. Rev. Lett.* **74** 3419; John S and Quang T 1994 *Phys. Rev. A* **50** 1764; Woldeyohannes M and John S 2003 *J. Opt. B: Quantum Semiclass. Opt.* **5** R43

[6] Joannopoulos J D, Villeneuve P R, and Fan S 1997 *Nature* **386** 143; Joannopoulos J D, Meade R D, and Winn J N 1995 *Photonic Crystal* (Princeton University Press, Princeton, NJ)

[7] Wang X-H, Gu B-Y, Wang R, and Xu H-Q 2003 *Phys. Rev. Lett.* **91** 113904; Wang X-H, Kivshar Y S, and Gu B-Y 2004 *Phys. Rev. Lett.* **93** 073901

[8] Cheng Y and Shen J Q 2007 ArXiv:gc-qc/0703066

[9] Hutton J T, Hannon J P and Trammell G T 1988 *Phys. Rev. A* **37** 4269; Hannon J P and Trammell G T 1999 *Hyperfine Interactions* **123/4** 127

[10] Baldwin G C and Solem J C 1997 *Rev. Mod. Phys.* **69** 1085; Baldwin G C and Solem J C, Gol'danskii V I 1981 *Rev. Mod. Phys.* **53** 687

[11] Prokhorov A M 1965 *Science* **149** 828; Brune M, Raimond J M, Goy P, Davidovich L, and Haroche S 1987 *Phys. Rev. Lett.* **59** 1899

[12] Batterman B W, Cole H, 1964 *Rev. Mod. Phys.* **36** 681

[13] Gerry C C and Knight P L 2005 *Introductory Quantum Optics* (CAMBRIDGE UNIVERSITY PRESS)

[14] Shih Y H 2003 *IEEE Journal of selected Topics in Quantum Electronics* **9** 1455

[15] Enaki N A 1988 JETP **67** 2033; Enaki N A and Macovei M 1997 *Phys. Rev. A* **56** 3274; Enaki N A and Mihalache D 1997 *Hyperfine Interactions* **107** 333

[16] Boschung B *et al.* 1995 *Phys. Rev. A* **51** 3650

[17] Fox M 2006 *Quantum Optics* (OXFORD UNIVERSITY PRESS)

[18] Josephson B D 1960 *Phys. Rev. Lett.* **4** 341; Pound R V and Rebka G A 1960 *Phys. Rev. Lett.* **4**, 274

[19] The strange data at $38^{th}$ min in the case III is due to acoustic shock induced by experimental quenching preparation.